\documentclass[pra,prl,floatfix,twocolumn, showpacs]{revtex4}
\usepackage{mathrsfs}
\usepackage{graphicx}
\usepackage{slashbox}
\usepackage{amssymb}
\usepackage{verbatim}

\begin{document}

\title{Irreducible multi-particle correlations in states without maximal rank}

\author{D.L. Zhou}
\affiliation{Beijing National Laboratory for Condensed Matter
Physics\\Institute of Physics, Chinese Academy of Sciences, Beijing
100080, China}

\begin{abstract}
In a system of $n$ quantum particles, the correlations are
classified into a series of irreducible $k$-particle correlations
($2\le k\le n$), where the irreducible $k$-particle correlation is
the correlation appearing in the states of $k$ particles but not
existing in the states of $k-1$ particles. A measure of the degree
of irreducible $k$-particle correlation is defined based on the
maximal entropy construction. By adopting a continuity approach, we
overcome the difficulties in calculating the degrees of irreducible
multi-particle correlations for the multi-particle states without
maximal rank. In particular, we obtain the degrees of irreducible
multi-particle correlations in the $n$-qubit stabilizer states and
the $n$-qubit generalized GHZ states, which reveals the distribution
of multi-particle correlations in these states.
\end{abstract}


\maketitle

\textit{Introduction.} --- How to classify and quantify different
types of correlations in a multi-particle quantum state is
fundamental in many-particle physics and quantum information. The
traditional method to characterize the multi-particle correlation in
many-particle physics is to use the multi-particle correlation
functions, which are directly associated with experimental
observables. Another method is originated from Shannon's deep
insight of information \cite{Shannon}, where the entropy is used as
a measure of information. Since different types of correlations in a
multi-particle state can be regarded as different types of nonlocal
information, a natural idea is to build a relation between
correlation and entropy.

Along this direction, the concept of the irreducible $n$-particle
correlation in an $n$-particle quantum state was first proposed in
Ref. \cite{Linden} by Linden \textit{et al.}. The counterpart for a
probability distribution of $n$ classical variables was given in
Ref. \cite{Schneidman}. Note that, the concept in Ref. \cite{Linden}
was naturally generalized in Ref. \cite{Schneidman}: a series of the
connected information of order $k$ are defined, which corresponds to
the irreducible $k$-particle correlation in an $n$-particle quantum
state.

The degree of irreducible $2$-particle correlation in a $2$-particle
quantum state is equal to the $2$-particle mutual entropy
\cite{Linden}, which is also obtained in different contexts
\cite{Groisman, Schumacher}. The irreducible $n$-particle
correlation in an $n$-particle state has been shown to be zero for
most $n$-particle pure states \cite{Linden2}, \textit{e.g.}, among
$n$-qubit pure states,  the irreducible $n$-particle correlation is
not zero only for the GHZ type pure states \cite{Walck}.

In this letter, based on the maximal entropy construction, we give
the definition of a measure of the degree of irreducible
$k$-particle correlation in an $n$-particle state. This definition
can be regarded not only as a direct generalization of the concept
in Ref. \cite{Linden}, but also as the quantum version of connected
correlation of order $k$ in Ref. \cite{Schneidman}. It is worthy to
note that the correlations in the $n$-particle system can be
classified into the irreducible $k$-particle correlations. In
another word, the degrees of irreducible multi-particle correlations
tell us how the correlations are distributed into the system.

Because the measure of the degree of irreducible $k$-particle
correlation is defined by the constrained optimization over the
$n$-particle quantum states, its explicit calculation in a general
$n$-particle state ($n>2$) is challenging, even for a $3$-qubit
state. To our best knowledge, no explicit calculations of
irreducible multi-particle correlations exist in the available
literature.

The main purpose of this letter is to provide a continuity approach
to calculate the degrees of irreducible multi-particle correlations
for the multi-particle states without maximal rank, which are
interested or useful in many particle physics or quantum
information. In particular, we obtain the analytic results for the
degrees of irreducible multi-particle correlations for the
stabilizer states \cite{Gottesman,Raussendorf,Hein} and the
generalized GHZ states \cite{Walck}.

\textit{Notations and definitions.} --- For simplicity, we introduce
the following notations. The set $\mathbf{e}(n)=\{1,2, \cdots, n\}$.
An $m$-element subset of the set $\mathbf{e} (n)$ is denoted as
$\mathbf{a}(m)=\{a_1, a_2, \cdots, a_m\}$, and the complementary set
of the set $\mathbf{a}(m)$ relative to the set $\mathbf{e}(n)$ is
denoted as $\mathbf{\bar{a}}(n-m)=\{\bar{a}_1, \bar{a}_2, \cdots,
\bar{a}_{(n-m)}\}$.

In a system of $n$ quantum particles, the complete knowledge of its
state is specified by the $n$-particle density matrix
$\rho^{\mathbf{e}(n)}$. The irreducible $k$-particle correlation
$(2\le k\le n)$ in the state is defined as the correlation appearing
in the $k$-particle reduced density matrixes $\rho^{\mathbf{a}(k)}$,
but not existing in the $(k-1)$-particle states
$\rho^{\mathbf{a}(k-1)}$. To define a measure of the degree of
irreducible multi-particle correlations in the state
$\rho^{\mathbf{e}(n)}$, similar to the method we adopted in Ref.
\cite{Zhou1}, we introduce an $n$-particle density matrix
$\tilde{\rho}_{l}^{\mathbf{e}(n)}$ for each $l\in \mathbf{e}(n)$
that satisfies:
\begin{equation}
\tilde{\rho}_l^{\mathbf{e}(n)}\in\{\rho_l^{\mathbf{e}(n)}|\mathrm{max}
S(\rho_l^{\mathbf{e}(n)})\},\label{eq1}
\end{equation}
where the $l$-particle reduced density matrix
\begin{equation}
\rho_l^{\mathbf{a}(l)}=\rho^{\mathbf{a}(l)} \label{eq2}
\end{equation}
for arbitrary subset $\mathbf{a}(l)$, and the function $S$ is the
von Neumann entropy defined as $S(x)=-\mathrm{Tr}(x \log_2 x)$. In
another word, the $n$-particle density matrix
$\tilde{\rho}_l^{\mathbf{e}(n)}$ has the same $l$-particle reduced
density matrixes as those of $\rho^{\mathbf{e}(n)}$, but it is
maximally noncommittal to the other missing information contained in
the state $\rho^{\mathbf{e}(n)}$ \cite{Jaynes}. A measure of the
degree of irreducible $k$-particle correlation in the state
$\rho^{\mathbf{e}(n)}$ is then defined as
\begin{equation}
C^{(k)}(\rho^{\mathbf{e}(n)})=S(\tilde{\rho}_{k-1}^{\mathbf{e}(n)})
-S(\tilde{\rho}_{k}^{\mathbf{e}(n)}). \label{cmk}
\end{equation}

The total correlation in the state $\rho^{\mathbf{e}(n)}$ is
referred to the nonlocal information appearing in
$\rho^{\mathbf{e}(n)}$, but not existing in the single particle
states $\rho^{\mathbf{a}(1)}$.  A measure of the degree of the total
correlation in the state $\rho^{\mathbf{e}(n)}$ is then defined as
\begin{equation}
C^{T}(\rho^{\mathbf{e}(n)})=S(\tilde{\rho}_{1}^{\mathbf{e}(n)})
-S(\tilde{\rho}_{n}^{\mathbf{e}(n)}). \label{cmt}
\end{equation}

Substituting Eqs. (\ref{cmk}) into Eq. (\ref{cmt}), we find that
\begin{equation}
C^{T}(\rho^{\mathbf{e}(n)})=\sum_{k=2}^{n}
C^{(k)}(\rho^{\mathbf{e}(n)}).\label{ct}
\end{equation}
Eq. (\ref{ct}) not only justifies that Eq. (\ref{cmt}) is a
legitimate measure of the total correlation, but also implies that
all the irreducible multi-particle correlations construct a
classification of the total correlation.

Note that the above definitions of the degrees of different types of
correlations, Eqs. (\ref{cmk}) and Eq. (\ref{cmt}), are intimately
related with the von Neaumann entropy. The underlying reason is as
follows. On the one hand, the von Neaumann entropy of a quantum
state is a measure of the degree of uncertainties of the state. On
the other hand, the existence of correlation in the multi-particle
quantum state decreases the uncertainties of the state. Therefore
the decreasing of uncertainties, \textit{i.e}, the entropy
difference, is reasonable to be used as a measure of the degree of
correlation.

\textit{Standard exponential form.} --- Note that the $n$-particle
density matrixes $\tilde{\rho}_l^{\mathbf{e}(n)}$ are essential
elements in the definitions in Eqs. (\ref{cmk}) and Eq. (\ref{cmt}).
The following important theorem gives the standard exponential form
of the state $\tilde{\rho}_l^{\mathbf{e}(n)}$.

\textbf{Theorem $1$:} For an $n$-particle quantum state
$\rho^{\mathbf{e}(n)}$ with maximal rank,  a state
$\tilde{\rho}_l^{\mathbf{e}(n)}$ ($1\le l\le n$) which satisfies
Eqs. (\ref{eq1}) and (\ref{eq2}) can be written in the following
exponential form:
\begin{equation}
\tilde{\rho}_l^{(\mathbf{e}(n))}=\exp \Big(\sum_{\mathbf{a}(l)}
\Lambda^{\mathbf{a}(l)}\otimes 1^{\mathbf{\bar{a}}(n-l)}\Big),
\label{ef}
\end{equation}
where $1^{\mathbf{\bar{a}}(n-l)}$ is the identity operators on the
Hilbert space of particles $\mathbf{\bar{a}}(n-l)$, and the
operators $\Lambda^{\mathbf{a}(l)}$ are to be determined by Eqs.
(\ref{eq2}).

\textbf{Proof:} The Lagrange multipliers $\Lambda^{\mathbf{a}(l)}$
are introduced to transform the constrained maximization defined by
Eqs. (\ref{eq1}) and (\ref{eq2}) to the following unconstrained
minimization:
\begin{eqnarray}
-S(\rho_m^{\mathbf{e}(n)})-\sum_{\mathbf{a}(l)}\mathrm{Tr}\big(\Lambda^{\mathbf{a}(l)}
(\rho_m^{\mathbf{a}(l)}-\rho^{\mathbf{a}(l)})\big)
\ge\mathrm{Tr}\big(\Lambda^{\mathbf{a}(l)}
\rho^{\mathbf{a}(l)}\big),\nonumber
\end{eqnarray}
where the Klein inequality \cite{Wehrl} is used. The equality is
satisfied if and only if Eq. (\ref{ef}) is satisfied. The Lagrange
multiplies $\Lambda^{\mathbf{a}(l)}$ are to be determined by Eqs.
(\ref{eq2}). Because the Klein inequality \cite{Wehrl} involves only
the positive operators, we need to limit ourselves to the states
with maximal rank.

A direct result of theorem $1$ is the following corollary.

\textbf{Corollary $1$:} For $m=1$ case, we have
\[
\tilde{\rho}_1^{\mathbf{e}(n)}=\exp \Big(\sum_{\mathbf{a}(1)}
\Lambda^{\mathbf{a}(1)}\otimes
1^{\mathbf{\bar{a}}(n-1)}\Big)=\prod_{i=1}^{n}\otimes\rho^{(i)}.
\]
Therefore the degree of the total correlation (\ref{ct}) in the
state $\rho^{\mathbf{e}(n)}$ is give by
\begin{equation}
C^{T}(\rho^{\mathbf{e}(n)})=\sum_{i=1}^{n}S(\rho^{(i)})
-S(\rho^{\mathbf{e}(n)}),\label{cte}
\end{equation}
where we used $\tilde{\rho}_n^{\mathbf{e}(n)}=\rho^{\mathbf{e}(n)}$.
Although the degree of the total correlation has an analytical
expression (\ref{cte}), we have not been able to give similar
analytical results for the degrees of irreducible multi-particle
correlations $C^{(k)}(\rho^{\mathbf{e}(n)})$.

Note that theorem $1$ is a direct generalization of the
corresponding result in Ref. \cite{Linden}. It shows that the
feature of multi-particle correlation in the state
$\tilde{\rho}_l^{\mathbf{e}(n)}$ is directly embodied in the
exponential form of the state. As emphasized in Ref. \cite{Linden},
theorem $1$ are not available for the multi-particle states without
maximal rank. However, most multi-particle states interested in many
particle physics or quantum information have not maximal rank,
\textit{e.g.}, the $n$-qubit stabilizer states and the generalized
GHZ states discussed below.

To overcome the difficulties in using theorem $1$ to treat with the
states without maximal rank, we adopt the following approach. A
multi-particle state without maximal rank can always be regarded as
the limit case of a series of multi-particle states with maximal
rank. If the degrees of irreducible multi-particle correlations for
the series of states with maximal rank can be obtained, then we take
the limit to get the degrees of irreducible multi-particle
correlations for the state without maximal rank. We called the above
method as the continuity approach. The proofs of theorem $2$ and $3$
below are typical applications of this approach.

\textit{Correlations in stabilizer states.} --- An $n$-qubit
stabilizer state state $\rho_{S}^{\mathbf{e}(n)}$ is defined as
\begin{equation}
\rho_{s}^{\mathbf{e}(n)}=\frac {1} {2^n}
\sum_{\{\alpha_i\in\mathbf{z}\}}\prod_{i=1}^{m} g_i^{\alpha_i},
\label{ss}
\end{equation}
where the set $\mathbf{z}=\{0,1\}$, $\{g_i\}$ are $m$ ($m\le n$)
independent commute $n$-qubit Pauli group elements. The set
$\mathfrak{g}(\rho_s^{\mathbf{e}(n)})=\{g_i\}$ is called the
stabilizer generator of the state $\rho_s^{\mathbf{e}(n)}$, and the
group generated by the generator $\mathfrak{g}$, denoted as
$\mathcal{G}(\rho_s^{\mathbf{e}(n)})=\{\prod_i g_i^{\alpha_i},
\alpha_i\in \mathbf{z}\}$, is called the stabilizer of the state. To
make the state (\ref{ss}) to be a legitimate state, the minus
identity operator is required not to be an element of the stabilizer
$\mathcal{G}$. Note that our definition of the $n$-qubit stabilizer
state is an extension of the usual definition \cite{Hein}, which
corresponds to the case when $m=n$. When $m<n$, the stabilizer
states defined by Eq. (\ref{ss}) are no longer pure states.

According to the definition of the $n$-qubit Pauli group, an element
$h\in \mathcal{G}(\rho_s^{\mathbf{e}(n)})$ can be written as $h=\pm
\prod_{i=1}^{n} O^{(i)}$ for $O\in \{I,X,Y,Z\}$, where $I$ is the
$2\times 2$ identity operator, and $X$, $Y$, $Z$ are three Pauli
matrixes. The number of identity operator $I$ in the element $h$ is
$N_I(h)=  \sum_i \mathrm{Tr} O^{(i)} /2 $. The stabilizer
$\mathcal{G}(\rho_{s}^{\mathbf{e}(n)})$ can be classified into a
series of sets $\mathcal{G}_k(\rho_{s}^{\mathbf{e}(n)})=\{h|h\in
\mathcal{G}(\rho_{s}^{\mathbf{e}(n)}), N_I(h)\ge n-k\}$ for $k\in
\mathbf{e}(n)$. Although in general
$\mathcal{G}_k(\rho_{s}^{\mathbf{e}(n)})$ is not a group, we can
still define the generator $\mathfrak{g}_k(\rho_s^{\mathbf{e}(n)})$
for the set $\mathcal{G}_k(\rho_{s}^{\mathbf{e}(n)})$ as a set of
elements in $\mathcal{G}_k(\rho_{s}^{\mathbf{e}(n)})$ such that
every element in $\mathcal{G}_k(\rho_{s}^{\mathbf{e}(n)})$ can be
written as a unique product of elements in the set. As in group
theory, the choice of the generator
$\mathfrak{g}_k(\rho_{s}^{\mathbf{e}(n)})$ is not unique in general.

The irreducible $k$-particle correlation in an $n$-qubit stabilizer
state is given by the following theorem.

\textbf{Theorem $2$:} The irreducible $k$-particle irreducible
correlation in an $n$-qubit stabilizer state
$\rho_{s}^{\mathbf{e}(n)}$ is
\begin{equation}
C^{(k)}(\rho_s^{\mathbf{e}(n)})=|\mathfrak{g}_k(\rho_s^{\mathbf{e}(n)})|
-|\mathfrak{g}_{k-1}(\rho_s^{\mathbf{e}(n)})|,
\end{equation}
where $|\cdot|$ denotes the size of a set.

\textbf{Proof:} Note that we can always take
$\mathfrak{g}_1(\rho_s^{\mathbf{e}(n)})\subseteq
\mathfrak{g}_2(\rho_s^{\mathbf{e}(n)}) \subseteq \cdots \subseteq
\mathfrak{g}_n(\rho_s^{\mathbf{e}(n)})$. Then the elements contained
in $\mathfrak{g}_{k}(\rho_s^{\mathbf{e}(n)})$ but not in
$\mathfrak{g}_{k-1}(\rho_s^{\mathbf{e}(n)})$ are reexpressed as
$g_{ki}$ for $ i\in
\mathbf{e}(|\mathfrak{g}_k|-|\mathfrak{g}_{k-1}|)$. Thus we can
construct an $n$-qubit state with a real parameter $\lambda$ as
\begin{equation}
\rho^{\mathbf{e}(n)}_{m}(\lambda)=\exp\Big(\eta+\lambda
\sum_{k=1}^m\sum_{i=1}^{|\mathfrak{g}_k|-|\mathfrak{g}_{k-1}|}g_{ki}\Big),
\end{equation}
where $\eta=-\ln( 2^n \cosh^{|\mathfrak{g}_m|}  \lambda)$, which is
determined by the normalization condition $\mathrm{Tr}
(\rho_{m}^{\mathbf{e}(n)})=1$. Then the above state is expanded as
\begin{equation}
\rho^{\mathbf{e}(n)}_{m}(\lambda)=\frac {1} {2^{n}}\Big(1
+\sum_{d=1}^{|\mathfrak{g}_m|}\tanh^d \lambda \sum_{\sum
\alpha_{ki}=d} \prod_{k\le m} g_{k i}^{\alpha_{k
i}}\Big).\label{eq11}
\end{equation}
Note that if $\exists$ $\alpha_{k i}=1$ for $k>m$, then $\forall
\mathbf{a} (m)$, we have  $\mathrm{Tr}_{\mathbf{\bar{a}}(n-m)}
\Big(\prod_{k\le n} g_{(k i)}^{\alpha_{k i}}\Big)=0$. Thus the
$m$-particle reduced density matrix $
\rho_m^{\mathbf{a}(m)}(\lambda)=\rho_n^{\mathbf{a}(m)}(\lambda)$.
According to theorem $1$, the degree of irreducible $k$-particle
correlation in the $n$-qubit state $\rho_n^{\mathbf{e}(n)}(\lambda)$
is
\begin{equation}
C^{(k)}(\rho_n^{\mathbf{e}(n)}(\lambda))=S(\rho_{k-1}^{\mathbf{e}(n)}
(\lambda))-S(\rho_{k}^{\mathbf{e}(n)}(\lambda)).
\end{equation}
From Eq. (\ref{eq11}), we observe that when the parameter $\lambda$
takes the limit of positive infinity, the states
$\rho^{\mathbf{e}(n)}_{m}(\lambda)$ are stabilizer states. In
particular,
\begin{equation}
\lim_{\lambda\rightarrow
+\infty}\rho^{\mathbf{e}(n)}_{n}(\lambda)=\rho_{s}^{\mathbf{e}(n)}.
\end{equation}
It is easy to prove that
$S(\rho_{m}^{\mathbf{e}(n)}(+\infty))=n-|l_m|$. Then the degree of
irreducible $k$-particle correlation in the stabilizer state
$\rho_{s}^{\mathbf{e}(n)}$ is
\begin{equation}
C^{(k)}(\rho_{s}^{\mathbf{e}(n)})=|\mathfrak{g}_k(\rho_{s}
^{\mathbf{e}(n)})|-|\mathfrak{g}_{k-1}(\rho_{s}^{\mathbf{e}(n)})|.
\end{equation}
Note that the result of theorem $2$ has been given in Ref.
\cite{Zhou}, which is based on some reasonable arguments in the
context of multi-party threshold secret sharing.

Theorem $2$ can be used to analyze the multi-particle correlation
distributions in all the stabilizer states. Let us illustrate its
power with analysis of the correlations in the two stabilizer
states: $\sigma_{1}^{\mathbf{e}(3)}=1/2(|000\rangle \langle 000|+
|111\rangle \langle 111|)$ and
$\sigma_{2}^{\mathbf{e}(3)}=|\textrm{GHZ}\rangle \langle
\textrm{GHZ}|$ with $|\textrm{GHZ}\rangle=1/\sqrt{2}(|000\rangle
+|111\rangle)$. A simple calculation yields the following results.
For the state $\sigma_{1}^{\mathbf{e}(3)}$, there are $2$ bits of
correlations altogether, and these $2$ bits of correlations are
irreducible $2$-particle correlations. For the other state
$\sigma_{2}^{\mathbf{e}(3)}$, the total correlations become $3$
bits, and these $3$ bits of correlations are classified into $2$
bits of irreducible $2$-particle correlation and $1$ bit of
irreducible $3$-particle correlation.

\textit{Correlations in generalized GHZ states.} --- A generalized
$n$-qubit GHZ states is defined as
\begin{equation}
|G_n\rangle =\alpha |00\cdots 0\rangle +\beta |11\cdots
1\rangle,\label{gs}
\end{equation}
where the parameters $\alpha$ and $\beta$ satisfy
$|\alpha|^2+|\beta|^2=1$ and $\alpha\beta \neq 0$. The degrees of
irreducible multi-particle correlations for the state
$\rho_G^{\mathbf{e}(n)}$ ($\equiv |G_n\rangle \langle G_n|$) are
given by the following theorem.

\textbf{Theorem $3$:} The degrees of irreducible multi-particle
correlation in the generalized GHZ state (\ref{gs}) are given by
\begin{eqnarray}
C^{(2)}(\rho_G^{\mathbf{e}(n)})&=&(n-1) E(|\alpha|^2),\\
C^{(n)}(\rho_G^{\mathbf{e}(n)})&=&E(|\alpha|^2),
\end{eqnarray}
where $E(x)=-x\log_2 x-(1-x)\log_2 (1-x)$. The degrees of the other
types of irreducible multi-particle correlation are zero
identically, \textit{i.e.},
$C^{(3)}(\rho_G^{\mathbf{e}(n)})=C^{(4)}(\rho_G^{\mathbf{e}(n)})=\cdots
=C^{(n-1)}(\rho_G^{\mathbf{e}(n)})=0$.

\textbf{Proof:} Let us construct an $n$-qubit state
\begin{equation}
\rho^{\mathbf{e}(n)}(\gamma,\vec{\lambda})= \exp\Big(\eta +\gamma
\sum_{i=2}^{n} Z^{(1)} Z^{(i)} +\vec{\lambda}\cdot \vec{\Sigma}
\Big), \label{ghzf}
\end{equation}
where the vector $\vec{\lambda}=\lambda_x \hat{x}+\lambda_y
\hat{y}+\lambda_z \hat{z}=\lambda \hat{\lambda}$ , the operator
vector $\vec{\Sigma}= \hat{x} X^{(1)}\prod_{i=2}^{n} X^{(i)}+ \hat
{y} Y^{(1)}\prod_{i=2}^{n} X^{(i)} +\hat{z} Z^{(1)} $, the parameter
$\eta$ is determined by the normalization condition
$\mathrm{Tr}\big(\rho^{\mathbf{e}(n)}(\gamma,\vec{\lambda})\big)=1$,
and the notation $\hat{v}$ represents the unit vector along the
direction of the vector $\vec{v}$. The $\hat{\lambda}$ component of
the operator vector $\vec{\Sigma}$ is denoted as
$\Sigma_{\lambda}=\hat{\lambda}\cdot \vec{\Sigma}$. Note that
$\Sigma_{\lambda}^{\dagger}=\Sigma_{\lambda}$,
$\Sigma_{\lambda}^2=1$, and $[\Sigma_{\lambda}, Z^{(1)}Z^{(i)}]=0$
for $i\in \mathbf{e}(n)$. The state (\ref{ghzf}) can thus  be
written as
$$
\rho^{\mathbf{e}(n)}(\gamma,\vec{\lambda})=\frac {1}
{2^{n}}\sum_{i=2}^{n}[1+\tanh(\gamma) Z^{(1)}
Z^{(i)}][1+\tanh(\lambda) \Sigma_{\lambda}].
$$
Note that in the above equation, only the term $Z_1\lambda_z/\lambda
$ in $\Sigma_{\lambda}$ contributes to the reduced $(n-1)$-particle
reduced density matrixes. Therefore the state
$\rho^{\mathbf{e}(n)}(\gamma,\vec{\lambda}^{\prime})$ has the same
$(n-1)$-particle reduced density matrixes as the state
$\rho^{\mathbf{e}(n)}(\gamma,\vec{\lambda})$ if the following
condition is satisfied:
\begin{eqnarray}
\lambda^{\prime}_x=\lambda^{\prime}_y=0,\;\;
\tanh\lambda^{\prime}_z=\frac {\lambda_z} {\lambda} \tanh\lambda.
\end{eqnarray}
According to theorem $1$, we find
\begin{equation}
\tilde{\rho}_m^{\mathbf{e}(n)}(\gamma,\vec{\lambda})=
\rho^{\mathbf{e}(n)}(\gamma,\vec{\lambda^{\prime}})
\end{equation}
for $m=2, 3, \cdots, n-1$. Therefore the degrees of irreducible
multi-particle correlations for the state
$\rho^{\mathbf{e}(n)}(\gamma,\vec{\lambda})$ can be obtained via
Eqs. (\ref{cmk}).

Without loss of generality,  we assume that in Eq. (\ref{gs})
$\alpha=\cos(\theta/2)$ and $\beta=\sin(\theta/2)e^{i\phi}$. Then we
define the Bloch vector
$\hat{u}=(\sin\theta\cos\phi,\sin\theta\sin\phi,\cos\theta)$. Let us
take $\hat{\lambda}=\hat{u}$, then $\Sigma_{u}
|G_n\rangle=|G_n\rangle$. The operators $\{Z^{(1)} X^{(i)}\}$ and
$\Sigma_u$ can be regarded as the stabilizer generator of the state
$\rho_G^{\mathbf{e}(n)}$. According to theorem $2$, the relation
between the generalized GHZ state $\rho_G^{\mathbf{e}(n)}$ and
$\rho^{\mathbf{e}(n)}(\gamma,\vec{\lambda})$ is
\begin{equation}
\rho_G^{\mathbf{e}(n)}=\lim_{\lambda\rightarrow
+\infty}\rho^{\mathbf{e}(n)}(\lambda,\lambda\hat{u}).
\end{equation}
In this case, we find that, for $m=2,3,\cdots, n-1$,
\begin{eqnarray}
&&\lim_{\lambda\rightarrow
+\infty}\tilde{\rho}_m^{\mathbf{e}(n)}(\lambda,\lambda\hat{u})\nonumber\\
&=&|\alpha|^2 |0 0\cdots 0\rangle \langle 0 0\cdots 0|+|\beta|^2 |1
1 \cdots 1\rangle \langle 1 1 \cdots 1 |.
\end{eqnarray}
A direct calculation will yield the results of theorem $3$.

Theorem $3$ shows that in the generalized $n$-qubit GHZ state
(\ref{gs}), only the irreducible $2$-particle correlation and the
irreducible $n$-particle correlation exist, and the previous one is
$(n-1)$ times of the later one.

\textit{Summary.} --- In summary, the definitions of the degrees of
irreducible $k$-particle correlations in an $n$-particle state are
given as a natural generalization of those defined in
\cite{Linden,Schneidman}. The significance of exponential form of a
multi-particle state in characterizing irreducible multi-particle
correlation is emphasized by theorem $1$. Adopting the continuity
approach, we are capable of applying theorem $1$ to deal with the
irreducible multi-particle correlations in the multi-particle states
without maximal rank. Particularly, we successfully obtained the
degrees of irreducible $k$-particle correlations in the $n$-qubit
stabilizer states and the $n$-qubit generalized GHZ states. The
multi-particle correlation structures in these states are revealed
by our results. We hope that the concepts of irreducible
multi-particle correlations will contribute to the characterizations
of multi-particle correlations in condensed matter system,
\textit{e.g.}, topological orders \cite{Kitaev,Wen,Yang} in
degenerate ground states.

The author thanks for helpful discussions with L. You, Z.D. Wang,
and C.P. Sun. This work is supported by NSF of China under Grant
10775176, and NKBRSF of China under Grant 2006CB921206 and
2006AA06Z104.


\begin{references}
\bibitem{Wehrl} A. Wehrl, Rev. Mod. Phys. \textbf{50}, 221 (1978).
The Klein inequality: for positive operators $A$ and $B$,
$\mathrm{Tr}(A(\ln A-\ln B))\ge 0$, the equality is satisfied if and
only if $A=B$.
\end{references}
\end{document}